**O. S. PASTOVEN**[1,2], Leading Engineer
https://orcid.org/0009-0000-4309-034X
**O. V. KOMPANIIETS**[1], Junior Researcher
https://orcid.org/0000-0002-8184-6520
E-mail: kompaniets@mao.kiev.ua
**I. B. VAVILOVA**[1], Dr. Sci. Hab. in Phys. Math., Prof., Head of Department
https://orcid.org/0000-0002-5343-1408
**I. O. IZVIEKOVA**[1,3], Junior Researcher
https://orcid.org/0009-0009-4307-0627

[1] Main Astronomical Observatory of the National Academy of Sciences of Ukraine
27, Akademik Zabolotnyi Str., Kyiv, 03143, Ukraine
[2] Faculty of Physics, Taras Shevchenko National University of Kyiv
4, Akademik Hlushkov Ave, Kyiv, 03680, Ukraine
[3] International Center for Astronomical, Medical and Ecological Research
27, Akademik Zabolotnyi Str., Kyiv, 03143, Ukraine

# NGC 3521 AS THE MILKY WAY ANALOGUE: SPECTRAL ENERGY DISTRIBUTION FROM UV TO RADIO AND PHOTOMETRIC VARIABILITY

*We studied the multiwavelength properties of NGC 3521, the Milky Way galaxy-twin, from UV- to radio, exploring the data from GALEX for UV-, SDSS for optical, 2MASS, WISE, MIPS (Spitzer) and PACS, SPIRE (Herschel) for IR-, and NRAO VLA for radio ranges. To obtain the spectral energy distribution (SED), we exploited the CIGALE software and constructed SEDs without (model A) and with (model B) AGN module. The type of nuclear activity of NGC 3521 is confirmed as the LINER.*

*We also present the results of the photometric data processing. Exploring the ZTF observations in 2018—2024, we found, for the first time, a weak photometric variability of the nuclear activity, where the correlation between g – r color indices and g-magnitude for long-term timescale shows a BWB trend (bluer-when-brighter) with a Pearson coefficient r(g – r) = 0.56, which is a medium correlation. To detect the variability of NGC 3521 during the day (IDV), we provided observations using a Zeiss-600 telescope with an aperture size of 8" at the Terskol observatory. The data obtained in the R-filter with an exposure of 90 sec for three hours on Feb 11, 2022, serve in favor of a trend towards an increase in brightness with the amplitude of variability of 0.04 ± 0.001 mag.*

*According to the results of the simulations, the best fit to the observed SED is provided by model A, which considers the contribution to the radiation from all galaxy components, assuming that the galaxy nucleus is inactive. Within this model, we derived the stellar mass $M_{star} = 2.13 \times 10^{10} M_{Sun}$, the dust mass $M_{dust} = 8.45 \times 10^7 M_{Sun}$, and the star formation rate $SFR = 1.76 M_{Sun}$/year with $\chi^2/d.o.f = 1.8$. Also, based on the HIPASS radio data, we estimated the mass of neutral hydrogen to be $M_{H\,I} = 1.3 \times 10^{10} M_{Sun}$, which is an order of magnitude greater than the mass of the stellar component.*

## 1. INTRODUCTION

The NGC 3521 is a nearby spiral flocculent galaxy at $z = 0.002672$. Its morphological type is SAB(rs) bc according to de Vaucouleurs classification, i.e., this galaxy has an intermediate bar structure, a weak inner ring, and moderately to loosely much-wound structure of spiral arms. A type of its nuclear activity is classified as H II LINER [15].

***Rotation curve.*** It is difficult to reveal a bar structure of NGC 3521 because of a high position angle of 163° and strong patchiness [54, 36] in the central region (Figure 1). The first estimations by Burbidge et al. in 1964 [7] allowed determining a) the mass from measures of the Hα and [N II] emission lines within about 170″ of the center as about $8 \times 10^{10}$ $M_{Sun}$, and b) specific rotation curve parameters. Later on, Casertano and van Gorkom [9] discovered this specificity of the rotation curve, which decreases with distance at the periphery. Dettmar and Skiff [16] in their photometric studies explored that NGC 3521 at low surface brightness levels has declining rotation curves with the typical signs of interaction or even merging. Zeilinger et al. [65] confirmed that NGC 3521 traces a specific counter-rotating population as well as exhibits asymmetric kinematic properties of gas and stars along its major and minor axes. Such a feature of NGC 3521 can be explained not only by the past merger but also in the frame of an axisymmetric dynamo concept, when density wave flows have spiral magnetic fields with a substantial radial component, amounting to 40—60 % of the azimuthal field in a differentially rotating flocculent galaxy [31].

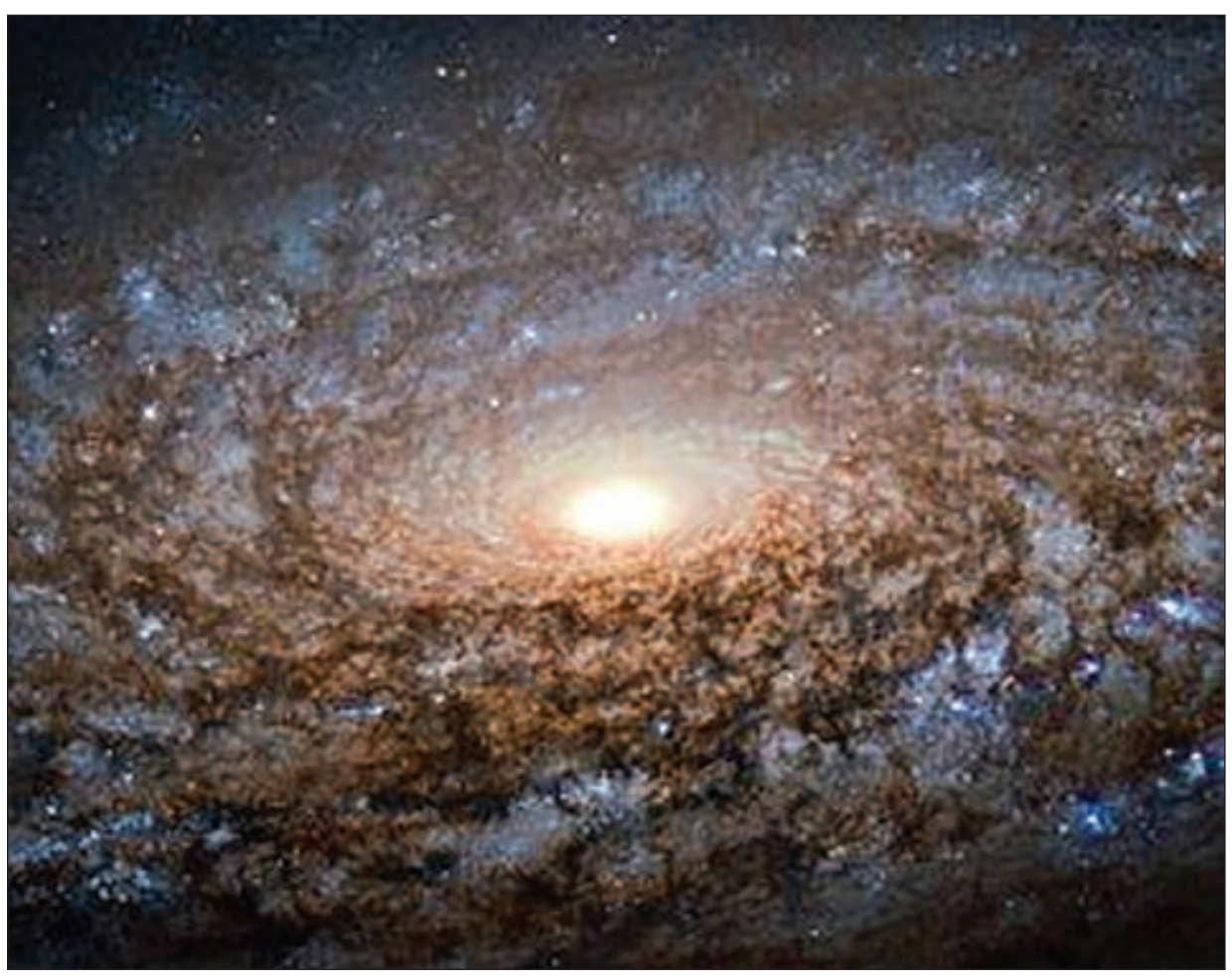

***Figure*** 1. A central part of the spiral barred galaxy NGC 3521 (the Bubble Galaxy) with extensive star-forming regions. Image is taken with the NASA/ESA Hubble Space Telescope in 2015 (file usage on Commons)

***Isolation criteria.*** NGC 3521 (= KIG 461) is an isolated galaxy (see, the Catalogue of Isolated Galaxies by Karachentseva [28]). Vavilova et al. [59] considered the isolation criterion as the first necessary condition to be a candidate for the Milky Way galaxy-analogue (MWA). The galaxy group around NGC 3521, together with the groups around NGC 3115 and NGC 2784, are part of a diffuse elongated Leo Spur structure in the large-scale web of the Local Volume [34]. At the same time, the properties of NGC 3521 and NGC 3115 have often been compared with those of the Milky Way and M 31, including their galaxies-satellites. Our Galaxy and M 31, more plausible, had a close flyby in the past 7—10 Gyr ago, and the study of their neighborhood brings clarity to understanding their evolution and hierarchical clustering. For instance, Haslbauer et al. [23] studied the Magellanic Clouds systems-analogues in the highest resolution TNG50 simulation using their stellar masses and distances in MW-like halos. They obtained that the Magellanic Clouds and their systems-analogues formed in physically unrelated ways in the ΛCDM model. Recently, investigating a neighboring group of galaxies around NGC 3521, which have six satellites, Karachentsev et al. [27] found four new likely satellites of low surface brightness in the projected area 750 × 750 kpc of the DECaLS survey (these objects have no yet the distance estimations). The total mass $M_T = (0.90 \pm 0.42)\ 10^{12}\ M_{Sun}$, the ratio $M_T/L_K = (7 \pm 3)\ M_{Sun}/L_{Sun}$, and peculiarities of the rotation curve [16] point out a shallow potential well and low mass of the dark halo of NGC 3521.

***Structural and multiwavelength properties.*** The isolated galaxies of the Local Volume exhibit a faint luminosity in spectral ranges, especially in radio- and X-ray [48, 56, 58, 60], and a weak nuclear activity as compared with galaxies in the dense environment [17, 40, 47, 62].

Below, we have collected several general results about NGC 3521 to picture the specific structural and multiwavelength features that this galaxy traces. Studying surface density profiles and size-mass relation, McGaugh [39] was the first to recognize

NGC 3521 as a Milky Way galaxy-twin based on the similarity of such parameters as the baryonic matter mass, rotation curve, and scale length of this galaxy. Subsequently, detailed studies of the spectral characteristics of NGC 3521 were performed by Pilyugin et al. [46]. By comparing the morphology, radial distributions of oxygen content, and rotation curves of the Milky Way and NGC 3521, these authors concluded that NGC 3521 is a twin of the Milky Way.

The most precise results on the structural properties in the optical range were obtained with the fiber-based Integral Field Unit instruments. For example, the high spectral resolution observations by Fabricius et al. [20] with the VIRUS-W not only confirmed a counter rotation of NGC 3521. Their investigations of the kinematics and metallicities [12] revealed a higher velocity dispersion in the bulge as well as a slower rotation compared to the disc. Identifying three main stellar populations by location and age, they proposed that the evolution of NGC 3521 began ≥7 Gyr (older stellar population), followed by a second burst of star formation or a merging about 3 Gyr ago when the bulge was firmed (intermediate stellar population), and undergone a new star formation burst in the disc ≤ 1 Gyr (young stellar population).

Asymmetries in the distributions of both stars and gas in the outermost regions are confirmed by the radio observational CO and H I data as well as the differences in the molecular gas properties and distribution along the major axis were mapped after processing the data of the Arizona Radio Observatory Survey [61]. Scaling relations for the giant molecular clouds in spiral arms and interarms based on the CO emission data in the PHANGS-ALMA survey with 90 pc resolution show that spiral arms clouds show slightly lower median virial parameters and their mass scale of 2.5× larger than interarm clouds [50]. The H II regions with enhanced and reduced chemical abundances are found distributed throughout the disc, and radial metallicity profiles have weak evidence of azimuthal variations [22].

NGC 3521 was included in several observational surveys, i.e., The H I Nearby Galaxy Survey of nearby galaxies (THINGS) in the H I line, which is based on data from the NRAO Very Large Array (VLA) [63]; the Nearby Galaxies Legacy Survey for the James Clerk Maxwell Telescope [64]; the Spitzer Infrared Nearby Galaxies Survey. It allowed determining the distance $d = 10.7$ Mpc by Tully-Fisher relation, radius $R = 4.16'$, isophotal radius $R_{25} = 12.94$ kpc. Using such a range of multiwavelength data, Warren et al. [64] examined the star formation properties, warm/dense molecular gas content, and dynamics on sub-kiloparsec scales.

Exploring the H I data from THINGS, Elson [19] found an anomalous diffuse and slow rotating H I component with $M_{\mathrm{H\,I}} = 1.5\times 10^9\ M_{Sun}$ (20 % of a total H I mass) and circular rotation speed, which is to lag the regular H I component by ~25—125 km/s. It is located in a thick disc (scale height ~3.5 kpc) and coincident with the inner regions of the stellar disc in the radial direction. This anomalous H I structure can be a slow-rotating halo gas component. The higher star formation rate in inner disc regions gives an explanation for a “galactic gas fountain” from the disc into the halo of NGC 3521. In turn, the radial profiles of stellar CO and 8 μm emission from polycyclic aromatic hydrocarbons (PAHs) obtained by Regan et al. [49] demonstrate that this galaxy has a central excess above the inward extrapolation of an exponential disc, in other words, the gas inflow rate into the central regions exceeds the star formation rate.

Among other multiwavelength properties of NGC 3521, we note the discovery of an ultrasonic X-ray source [SST2011] J110545.62 + 000016.2 by Heida et al. [24]. It was later analyzed by Lopez et al. [37] with VLT X-shooter and Chandra data. They found a connection of the ULX source with an adjacent H II region at ~138 pc to the NE and obtained its X-ray luminosity in the 0.3—7 keV range as $(1.9 \pm 0.8) \times 10^{40}$ erg/s cm$^2$.

We chose this galaxy for a detailed multiwavelength analysis to determine its main characteristics by modelling the spectral energy distribution (SED) and accumulate the data on how the SEDs of galaxies can be indicators for the search of the Milky Way galaxies-analogues (MWAs).

***The structure of our article*** is as follows. We describe the data from sky surveys in Section 2 and the main modules of the CIGALE software, which we used for the multiwavelength analysis, in Section 3. The results for the spectral energy distribution models with and without the AGN module are presented in Sec-tion 4, as well as on the photometric studies obtained with the Zeiss-600 telescope at the Terskol observatory in 2021—2022 and with the Zwicky Transient Facility in 2018—2024 in Section 5. A summary of the results with a brief discussion is given in Section 6.

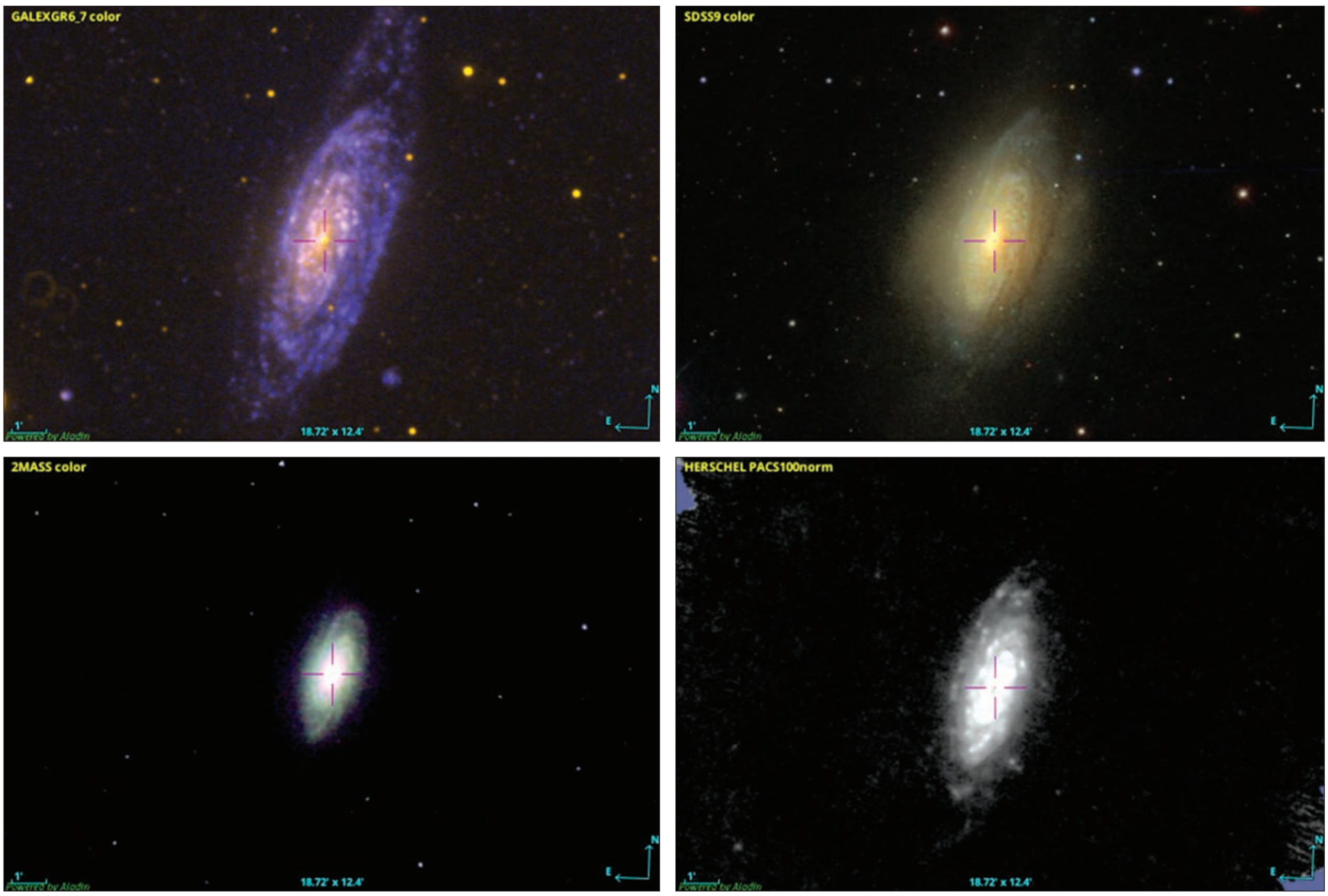

***Figure* 2.** NGC 3521 in the ultraviolet, visible, and infrared ranges. The images were taken with the Aladin software [3]

## 2. THE MULTIWAVELENGTH DATA

We used the data from various publicly available sky surveys for the multiwavelength analysis. Most of them are collected in the NED (NASA/IPAC Extragalactic Database). To cover the different bands, we used GALEX for ultraviolet (UV), SDSS for optical, 2MASS, WISE, MIPS (Spitzer) and PACS, SPIRE (Herschel) for infrared (IR), and NRAO VLA for radio data. Altogether they formed a multiwavelength range from 135 nm to 21 cm.

NGC 3521 is a nearby large galaxy. So, the standard apertures for which the fluxes are calculated in these surveys are much smaller than the galaxy size. For this reason, in our work, we used the fluxes obtained from the aperture of 263″ by 56″ from the Spectral Energy Distribution atlas for 129 nearby galaxies [5] in the range from UV to near-IR. The data from the Herschel telescope (PACS and SPIRE) were taken from the work [14]: the region from which the fluxes were obtained was an ellipse with axes 926″ by 455″. Observational fluxes from the Spitzer telescope were obtained for the entire visible part of the galaxy [43] as well as the radio data from [13].

## 3. MAIN MODULES IN THE CIGALE SOFTWARE

The flexibility of the CIGALE software [4] allows working with the data from various sky surveys and telescopes. As part of the CIGALE data processing, we consider various components of the galaxy's radiation: stellar population, dust and gas emission, active galactic nuclei. To do this, we used the following independent modules.

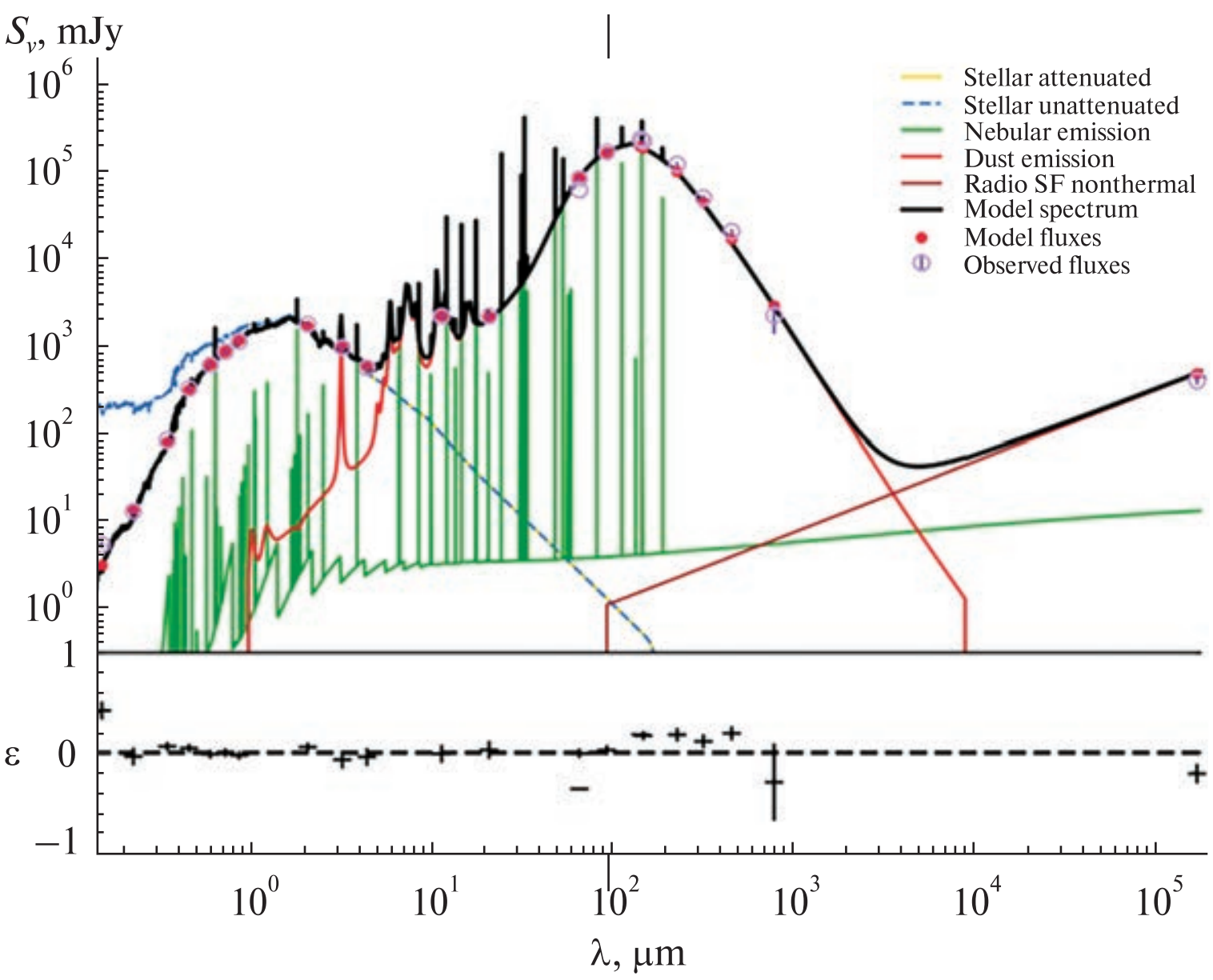

***Figure*** 3. NGC 3521. Best-fit SED from the UV to radio ranges using model A. Red dots are the model fluxes; purple dots are the observed fluxes. The green colour describes the nebular emission; the yellow colour describes the stellar component without attenuation; the blue dashed lines describe the stellar component with attenuation; the red colour describes the dust emission; and the brown colour describes the synchrotron emission from star-forming regions. The black continuous curve is the resulting model SED

The **sfhdelayedbq** module represents the model for describing the history of star formation in the galaxy — the so-called modified, delayed star formation rate model with an optional exponential explosion [11]. It allows us to consider the recent decay of the star formation rate as

$$SFR(t) \propto \begin{cases} t \times \exp(-t/\tau) & t \le t_0, \\ r_{SFR} \times SFR(t = t_0) & t > t_0, \end{cases} \quad (1)$$

where $t_0$ is the time of the rapid decrease in the $SFR$, $\tau$ is the time when the $SFR$ reaches its maximum. The parameter $r_{SFR}$ in Eq. (1) is the ratio between $SFR$ ($t > t_0$) and $SFR$ ($t = t_0$):

$$r_{SFR} = \frac{SFR(t>t_0)}{SFR(t=t_0)}. \quad (2)$$

The $SFR$ (Eq. (1)) is determined by four parameters: the age of the stellar population, the exponential decay time of the model, the star formation time of the main stellar population, and the age of the outburst or extinction episode, $r_{SFR}$ (Eq. (2)).

The stellar emission is computed using the **bc03** stellar population synthesis [6]. This library covers a wide range of metallicity values (0.0001, 0.0004, 0.004, 0.008, 0.02, and 0.05). The input parameters for calculating the stellar population spectrum — the metallicity of the stars and the mass function describing the different components of the galaxy — were considered in [10]. Massive OB stars ionize the gas around them. This gas re-emits energy through emission lines and a continuum (caused by free-free, free-bound, and two-photon transitions). This radiation is important for studying recent star formation through the hydrogen lines and radio continuum and the metallicity of the gas through the metal lines. We used the nebular module based on nebula templates from [26] to model the nebular emission. They are parameterized by the ionization coefficient $U$ and the gas metallicity $Z$. The electron density is assumed to be constant and equal to 100 cm$^{-3}$. The ionization parameter is the ratio of the flux of ionized photons to the density of hydrogen atoms [44] and is expressed as Eq. (3):

$$U = \frac{ionizing\ photon\ flux}{cn_e} = \frac{1}{4\pi r^2}\int_{\nu_0}^{\infty} L_\nu \frac{d\nu}{h\nu}. \quad (3)$$

Dust in galaxies absorbs shortwave radiation very efficiently. The energy absorbed from the UV to the near-IR is re-emitted in the mid- and far-IR.The **dustatt_modified_starburst** module is based on the starburst decay curve obtained by Calzetti et al. [8] and extended by Noll et al. [45] to add dust features in the vicinity of 220 nm. The resulting expression of the corresponding attenuation coefficient is given in Eq. (4), where the change in flow as a result of interaction with the medium is taken into account:

$$F_{intr}\left(\lambda\right) = F_{obz}\left(\lambda\right)\cdot 10^{0.4 E_l\left(B-V\right)k_\lambda}, \qquad (4)$$

where $F_{intr}(\lambda)$ is the absorption-corrected flux, $F_{obz}(\lambda)$ is the observational flux, $E_l(B-V) = f\cdot E(V_B)$ is the reddening in the line, which is obtained by multiplying the stellar extinction by the coefficient $f$. The standard value is $f = 0.44$.

$$k_\lambda = D_\lambda + k_\lambda^{starbars}\cdot(\lambda/550\,\mathrm{nm})\cdot\frac{E\left(B-V\right)_{\delta=0}}{E\left(B-V\right)_\delta}, \qquad (5)$$

$$D_\lambda = \frac{E_{bump}\lambda^2\gamma^2}{(\lambda^2-\lambda_0^2)^2+\lambda^2\gamma^2},$$

where the latter expression for $D_\lambda$ is the DRUID profile, which characterizes the UV bump in the galaxy spectrum. The DRUID profile depends on the wavelength corresponding to the maximum intensity ($\lambda_0$), FWHM ($\gamma$), and amplitude ($E_{bump}$). The nature of the bump is still unknown. $k_\lambda^{starbars}$ is the effective stellar continuum attenuation curve normalized by the gas extinction coefficient. $\delta$ is the slope of the UV continuum. For dust clouds around young stars, in general, $\delta_{BC} = -1.3$, for the interstellar medium $\delta_{ISM} = -0.7$. The dust cloud and the interstellar medium attenuate radiation from stars younger than 10 Myr; stars that are older than 10 Myr attenuate only in the interstellar medium.

Cosmic dust not only attenuates but can also emit radiation [18]. The module describing this radiation is **dl2014**. This module is based on a study of radiation from a mixture of amorphous silicate dust, graphite grains, and PAHs. One of the critical features of this module is the division of dust emissions into two components. The first model is the diffuse radiation of dust heated by the total stellar population. Here, the dust is illuminated by a single radiation field $U_{\min}$. The second model simulates dust closely associated with star-forming regions. In this case, the dust is illuminated by a variable radiation field in the range from $U_{\min}$ to $U_{\max}$ and is given by Eq. (6):

$$\frac{dM_{dust}}{dU} = \left(1-\gamma\right)M_{dust}\delta\left(U-U_{\min}\right) + \gamma M_{dust}\frac{\left(\alpha-1\right)}{\left(U_{\min}^{-\alpha}-U_{\max}^{-\alpha}\right)}U^{-\alpha}, \qquad (6)$$

where $dM_{dust}$ is the mass of dust heated by radiation in the interval $[U,\ U + dU\,]$, $M_{dust}$ is the total mass of dust, $(1 - \gamma)$ is the fraction of the dust mass that is exposed to the starlight intensity $U_{\min}$, and $\alpha$ is the power law exponent. In the model, $U_{\max} = 10^7$. The last parameter of this model is $q_{PAH}$, the mass fraction of surfactants common to the two components.

Radiation in the radio range is described using the **radio** module, which is implemented in the CIGALE software [4]. The model can include the synchrotron component's contribution from the AGN and star formation regions. Due to a large number of parameters, the radio module uses the correlation between the radio and infrared bands $q_{IR}$, which was proposed in [25], the power spectral slope $\alpha$, and the assumption that at a wavelength of 21 cm, the spectrum is mainly dominated by non-thermal radiation. The correlation coefficient $q_{IR}$ is defined as $q_{IR} = (F_{FIR}/3.75\times10^{12}\ \mathrm{Hz})\ /\ S_p(1.4\ \mathrm{GHz})$, where $F_{FIR} = 1.26\times10^{-14}\times[2.58 f_\nu(60\ \mu\mathrm{m}) + f_\nu(100\ \mu\mathrm{m})] f_\nu$ is given in Jy, and $F_{FIR}$ in W m$^{-2}$; $3.75\times10^{12}$ Hz is the frequency value at a wavelength of 80 μm. Given the modelled IR data, the correlation parameter $q_{IR}$ is used to estimate the flux at a wavelength of 21 cm. On the other hand, the radio data can help estimate IR radiation if no other data are available in this range.

To describe the active nucleus, we used the **fritz2006** module, which is based on the model by Fritz et al. [21]. It explicitly considers three components of radiation: from a source located n the torus, scattered by dust, and thermal radiation from dust. This module is determined by a set of seven parameters: $r_{radio}$ is the ratio of the maximum to the minimum radii of the dust torus, $\tau$ is the optical thickness at 9.7 μm, $\beta$ and $\gamma$ describe the dust density distribution ($\propto r^\beta\cdot e^{-\gamma|\cos\theta|}$), where $r$ is the radius, and $\theta$ is the opening angle of the dust torus, $\psi$ is the angle between the axis of the AGN and the visual beam, and the fraction of the AGN $frac_{AGN}$. Also, the redshifting module was applied to all the models to account for the redshift.

*Table 1.* **NGC 3521. Input parameters of the SED model A**

| Parameter | Values |
|---|---|
| Star formation history, **sfhdelayedbq** | |
| E-folding time of the main stellar population model, in Myr | 500, 1000.0, 2000.0, 5000 |
| Age of the main stellar population in the galaxy, in Myr | 5000, 10000, 13000 |
| Age of the burst/quench episode, in Myr | 10, 25, 50, 100 |
| The ratio of the *SFR* after/before age_bq | 0.001, 0.01, 0.1,0.5, 1, 2.5, 5, 10, 50 |
| Stellar population, **bc03** | |
| Initial mass function | [18] |
| Metallicity | 0.02, 0.05 |
| Age in Myr of the separation between the young and the old star populations | 10 |
| Nebular emission, **nebular** | |
| Ionisation parameter | –3.0 |
| Gas Metallicity | 0.02, 0.051 |
| Dust attenuation, **dustatt_modified_starburst** | |
| $E_l(B-V)$, the colour excess of the nebular lines light for both the young and old population | 0.3, 0.5, 0.75, 1 |
| Slope delta of the power law modifying the attenuation curve | –1.3, –0.7, 0.0 |
| Dust emission, **dl2014** | |
| Mass fraction of PAH | 0.47, 1.12, 2.5, 3.90, 5.95 |
| Minimum radiation field | 2, 5, 10, 25 |
| Power-law slope $dU/dM \propto U^{\alpha}$ | 1, 2, 3 |
| Fraction illuminated from $U_{\min}$ to $U_{\max}$ | 0.1, 0.25, 0.5 |
| Synchrotron radio emission, **radio** | |
| The value of the FIR/radio correlation coefficient for star formation | 2.58 |
| The slope of the power-law synchrotron emission is related to star formation | 0.8 |

## 4. THE SPECTRAL ENERGY DISTRIBUTION MODELS FOR NGC 3521

**4.1.** ***SED from UV to radio ranges without taking into account the AGN module.*** The best statistical agreement between the observed and theoretical energy distributions was obtained with a value of $\chi^2/d.o.f =$ $= 1.8$. The resulting values of the stellar mass $M_{star} =$ $= 2.13\times10^{10}\ M_{Sun}$, dust mass $M_{dust} = 8.45\times10^7\ M_{Sun}$, and star formation rate $SFR = 1.76\ M_{Sun}$/year. In this model, the star formation rate is the closest to the values obtained in other studies. We will discuss this in more detail in the next section. We also note that we have selected the best value of the ionization parameter $\lg U = -3$ from those listed in Table 1.

Since none of the models used allows us to estimate the gas mass from the available observational data, we used the equation obtained by van Gorkom [55] to derive the observed mass of H I, which is also recommended for use on the VLA portal[1]:

$$M_{\rm H\,I} = 2.36\cdot10^5 D^2 S\Delta V\left[M_{Sun}\right], \qquad (7)$$

where $M_{\rm H\,I}$ is the observed mass of neutral hydrogen, $D$ is the distance to the galaxy, and $S\Delta V$ is the H I line area in units of Jy km/s. We took the value of $S\Delta V$ from the catalogue [1]. Thus, the mass of H I for this galaxy, according to the observational data in the radio range, is $M_{\rm H\,I} = 1.3\times10^{10}\ M_{Sun}$, which is of an order of magnitude larger than the mass of the stellar component.

**4.2.** ***SED from UV to radio ranges with taking into account the module of AGN.*** NGC 3521 was classified as a galaxy with an active nucleus of the LINER type using IR observational data [53]. The same classification is reported in several other studies, see, for example, [22]. However, in the work by Serote Roos et al. [52] this classification was not confirmed based on the spectra obtained in the optical range for the central part. At the same time, Moustakas and Kennicutt [41] performed several spectrophotometric observations in the optical range for different regions of NGC 3521 (see, Figs 4 and 5). To clarify the activity class of NGC 3521, we constructed the Baldwin-Phillips-Terlevich diagram [2] The relation for this separation was derived by Kewley et al. [30], where a combination of photoionization and stellar population synthesis models was used to establish a theoretical upper limit on the location of galaxies with star formation outbursts in BPT diagrams:

$$\lg\left(\left[\mathrm{O\ III}\right]\left(500.7\,\mathrm{nm}\right)/\mathrm{H}_\beta\right) = \frac{0.61}{\lg\left(\left[\mathrm{N\ II}\right]\left(658.5\,\mathrm{nm}\right)/\mathrm{H}_\alpha\right) - 0.47} + 1.19. \qquad (8)$$

[1] https://science.nrao.edu/facilities/vla/docs/manuals/oss2013B/performance/referencemanual-all-pages

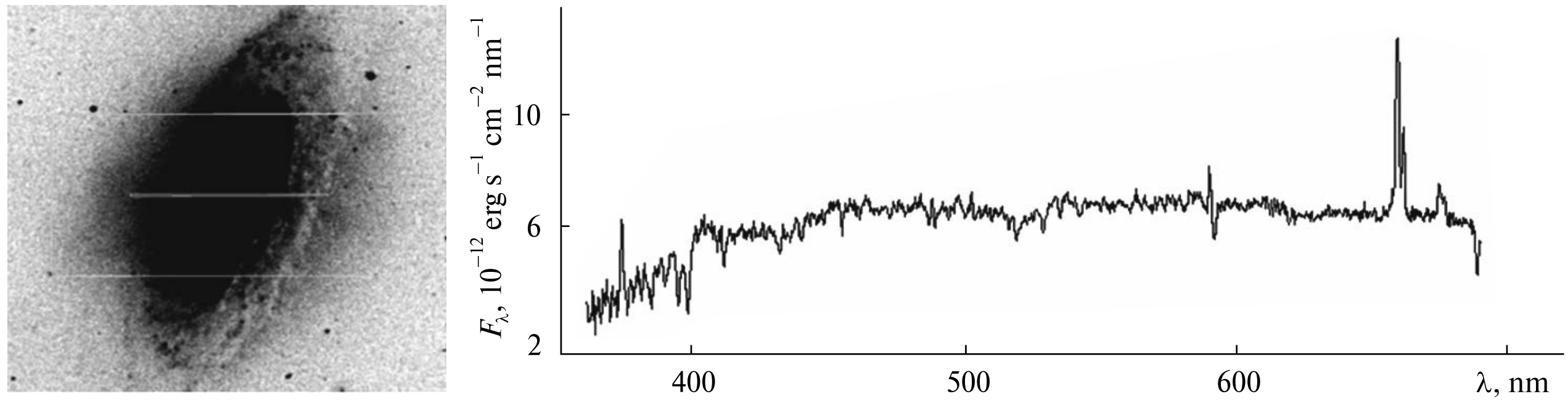

***Figure*** 4. NGC 3521. The spectrum of the galaxy for a region of 190 by 360 angular seconds was obtained using the Boller-Chivens spectrograph on the 2.3-meter Bok telescope at Kitt Peak Observatory [41]

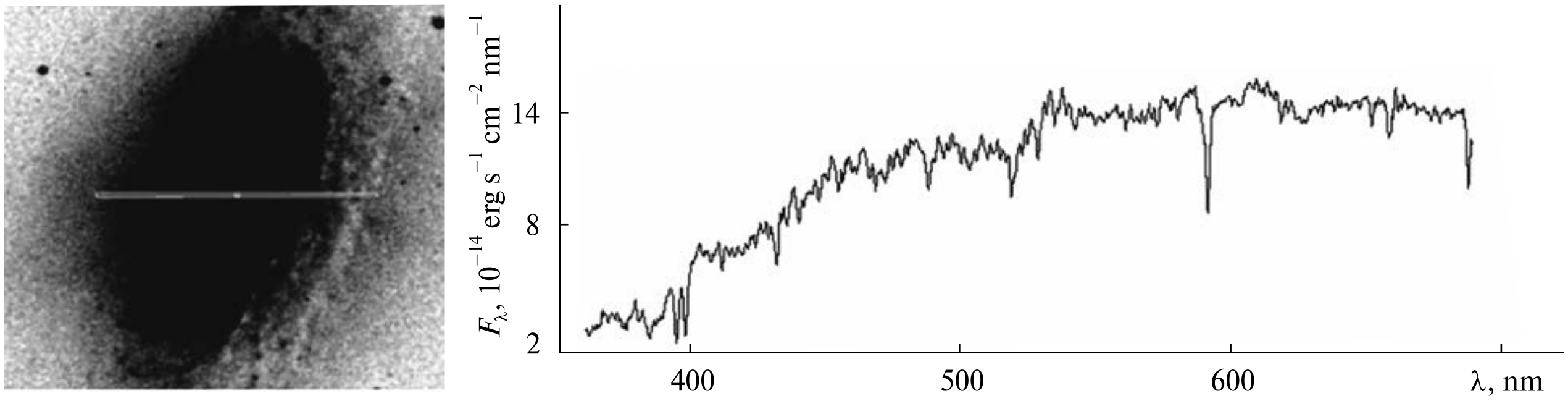

***Figure*** 5. NGC 3521. The spectrum of the region close to the nucleus for an area of 2.50 by 2.50 angular seconds was obtained using the Boller-Chivens spectrograph on the 2.3-meter Bok telescope at Kitt Peak Observatory [41]

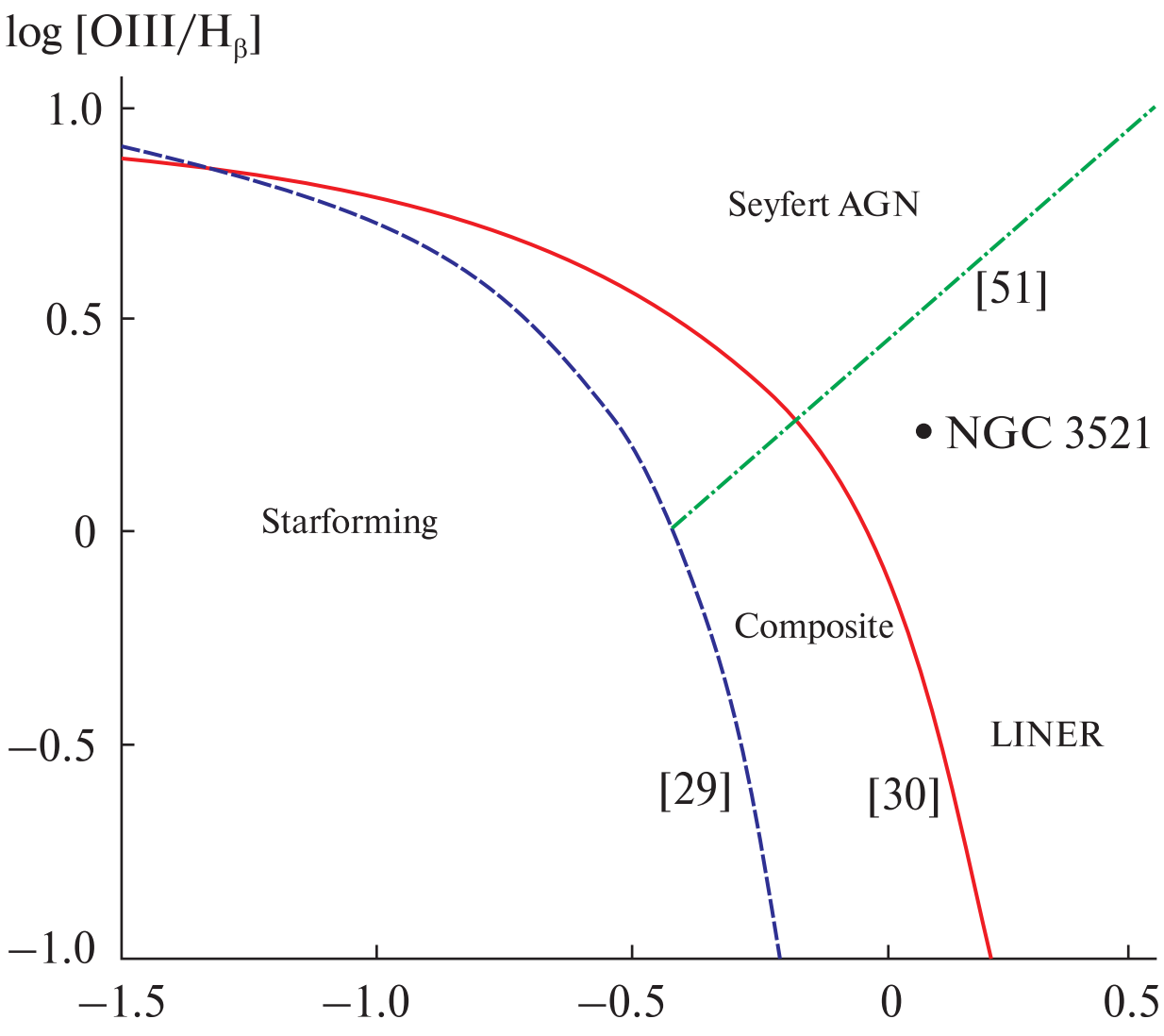

***Figure*** 6. BPT diagram with values for the galaxy NGC 3521. The red line is the upper limit of galaxies with active star formation (curve 8), the blue dashed line is the active star formation partition curve (curve 9), the green dashed line is the empirical ratio for the Seyfert-LINER partition (curve 10).

An empirical relation for the distinction of AGNs is introduced by Kauffmann et al. [29]:

$$\lg([\mathrm{O\ III}](500.7\,\mathrm{nm})/\mathrm{H}_\beta) = \frac{0.61}{\lg([\mathrm{N\ II}](658.5\,\mathrm{nm})/\mathrm{H}_\alpha) - 0.05} + 1.3. \quad (9)$$

In Figure 6, the area between curves 8 and 9 is the region of the so-called composites, i. e., those galaxies that have the characteristic radiation of both star-forming galaxy and galaxy with active nucleous. The empirical division of active galaxies into Seyfert and LINERs is introduced by Schawinski et al. [51] and presented in Eq. (10):

$$\lg([\mathrm{O\ III}](500.7\,\mathrm{nm})/\mathrm{H}_\beta) = 1.05 \cdot \lg([\mathrm{N\ II}](658.5\,\mathrm{nm})/\mathrm{H}_\alpha) + 0.45. \quad (10)$$

To determine the class of NGC 3521, we used the calculated fluxes from the catalogue [42] (the data taken from the table6.dat "Nuclear emission-line fluxes"). Having the flux ratio, we constructed a BPT diagram for NGC 3521. The resulting location of NGC 3521 on the BPT diagram based on the above

*Table 2.* **NGC 3521. Input parameters of the SED model B (with AGN module)**

| Parameter | Values |
|---|---|
| Star formation history, **sfhdelayedbq** | |
| E-folding time of the main stellar population model, in Myr | 500, 1000.0 |
| Age of the main stellar population in the galaxy, in Myr | 7000, 10000 |
| Age of the burst/quench episode, in Myr | 10, 25, 50, 100 |
| The ratio of the *SFR* after/before age_bq | 0.001, 0.01, 0.1,0.5, 1, 2.5, 5, 10 |
| Stellar population, **bc03** | |
| Initial mass function | Chabrier [18] |
| Metallicity | 0.02, 0.05 |
| Age in Myr of the separation between the young and the old star populations | 10 |
| Nebular emission, **nebular** | |
| Ionisation parameter | –3.0 |
| Gas Metallicity | 0.0004, 0.02 |
| Dust attenuation, **dustatt_modified_starburst** | |
| $E_l(B - V)$, the colour excess of the nebular lines light for both the young and old population | 0.5, 0.75, 1, 1.6 |
| Slope delta of the power law modifying the attenuation curve | –1.3, –0.7, 0 |
| Dust emission, **dl2014** | |
| Mass fraction of PAH | 0.47, 1.12, 2.5, 3.90, 5.95 |
| Minimum radiation field | 2, 5, 10, 25 |
| Power-law slope $dU/dM \propto U^{\alpha}$ | 1, 2, 3 |
| Fraction illuminated from $U_{min}$ to $U_{max}$ | 0.1, 0.25, 0.5 |
| AGN, **fritz2006** | |
| The ratio of the maximum to minimum radii of the dust torus | 60 |
| The full opening angle of the dust torus | 60 |
| The angle between the equatorial axis and line of sight | 0.001, 89.99 |
| AGN fraction | 0.01 |
| $E(B\text{-}V)$ for the extinction in the polar direction in magnitudes | 0.25, 0.5, 0.7 |
| Synchrotron radio emission, **radio** | |
| The value of the FIR/radio correlation coefficient for star formation | 2.58 |
| The slope of the power-law synchrotron emission is related to star formation | 0.8 |
| The radio-loudness parameter for AGN | 5 |
| The slope of the power-law AGN radio emission (assumed isotropic) | 0.7 |

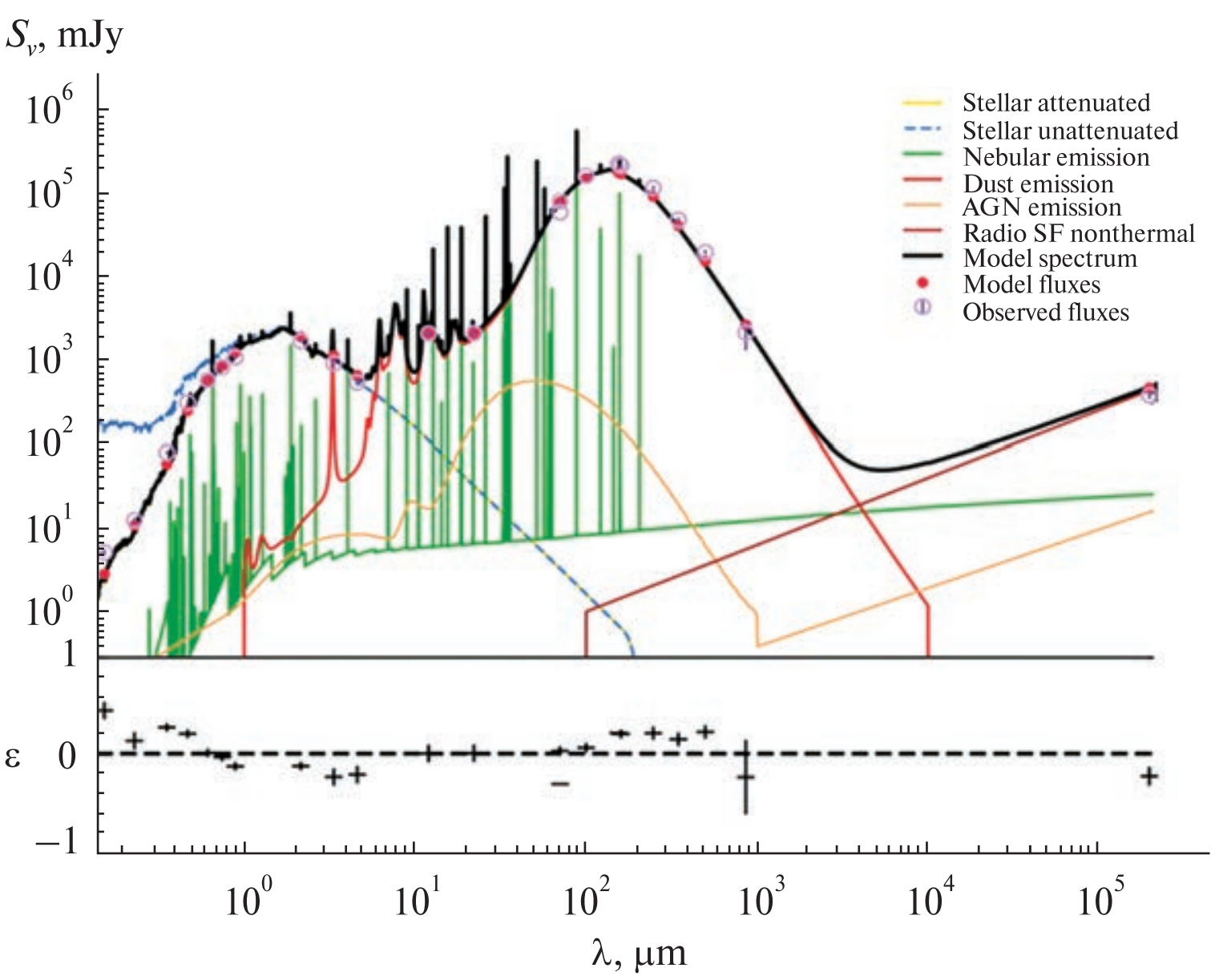

***Figure*** 7. The best approximation of the SED from UV to radio ranges using model B. Red dots are model fluxes, and purple dots are observed fluxes. The green colour describes the nebular radiation, the yellow colour describes the stellar component without attenuation, the blue dashed lines describe the stellar component with attenuation, the orange colour describes the radiation from the AGN, the red colour describes the dust radiation, and the brown colour describes the synchrotron radiation from the star formation regions and the AGN. The black continuous curve is the resulting model SED

observations allows us to classify it as a LINER-type galaxy with an active nucleus (Fig. 6).

Since we do not know the characteristics of the AGN, we consider various variations as input parameters in the **fritz2006** model, including the angle at which we observe the active nucleus. It is also worth noting that the computational capabilities of the computer limit the number of possible combinations for each module. We explored the same approach as for 18 2MIG isolated galaxies with active nuclei, which were modeled with CIGALE [32].

Most of these parameters do not vary; we changed their values during different test runs, and here we present the combination that gives us the best statistical result with $\chi^2/d.o.f = 2.8$ for the model with the AGN. The corresponding mass of dust is $M_{dust} = 8.05 \times \times 10^7\ M_{Sun}$, the mass of stars $M_{star} = 3.34 \times 10^{10}\ M_{Sun}$, the star formation rate $SFR = 2.17\ M_{Sun}$/year, and the contribution of the AGN to the total radiation $frac_{AGN} = 0.01$. Compared to the results obtained with model A, the model B with AGN module led to a slight increase in $\chi^2/d.o.f.$ The parameters obtained in both models were similar in value and order.

## 5. PHOTOMETRIC VARIABILITY OF NGC 3521 BY THE DATA WITH ZWICKY TRANSIENT FACILITY AND THE ZEISS-600 TELESCOPE AT THE TERSKOL OBSERVATORY

Due to the close location of NGC 3521 to the celestial equator, it is visible for a short time from both hemispheres mostly at the beginning of the year that complicates regular observations and monitoring of activity of its nucleus.

***Zeiss-600 telescope data.*** Nevertheless, for revealing and studying the photometric variability of its active nucleus we included it in the observational program with the Zeiss-600 telescope with an 8  aperture at the Terskol Observatory in 2021—2022. The data processing was carried out using the MaximDL software package. Table 3 summarizes the data about the variability of NGC 3521, where  is the variability amplitude, and *Er*± is the largest error in the given filter; the light curve is shown in Figure 8.

In order to detect the variability of NGC3521 during the day (IDV), several observations were carried out with the help of the Zeiss-600 telescope. Namely, observation was made in the R-filter with an expo-sure of 90 seconds for three hours on Feb 11, 2022. We can see in Fig. 8 (right) that there is a trend towards an increase in brightness during three hours, the amplitude of variability during observation was $0.04 \pm 0.001$ mag.

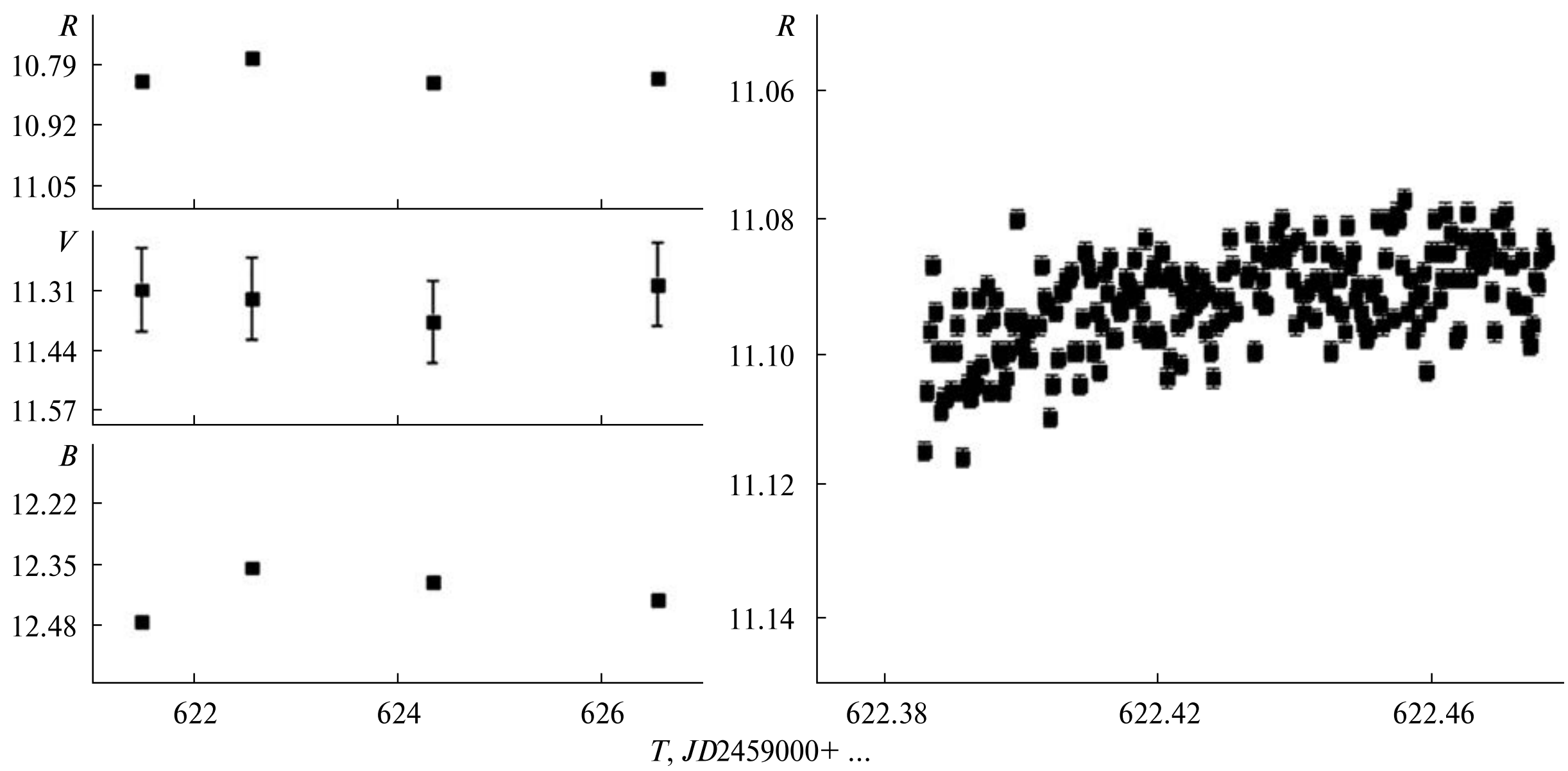

***Figure*** 8. Light curves for NGC 3521 as observed at the Terskol Observatory in 2021—2022: in *BVR* filters (left), in the *R* filter (right) over 3 hours of observation on Feb 11, 2022

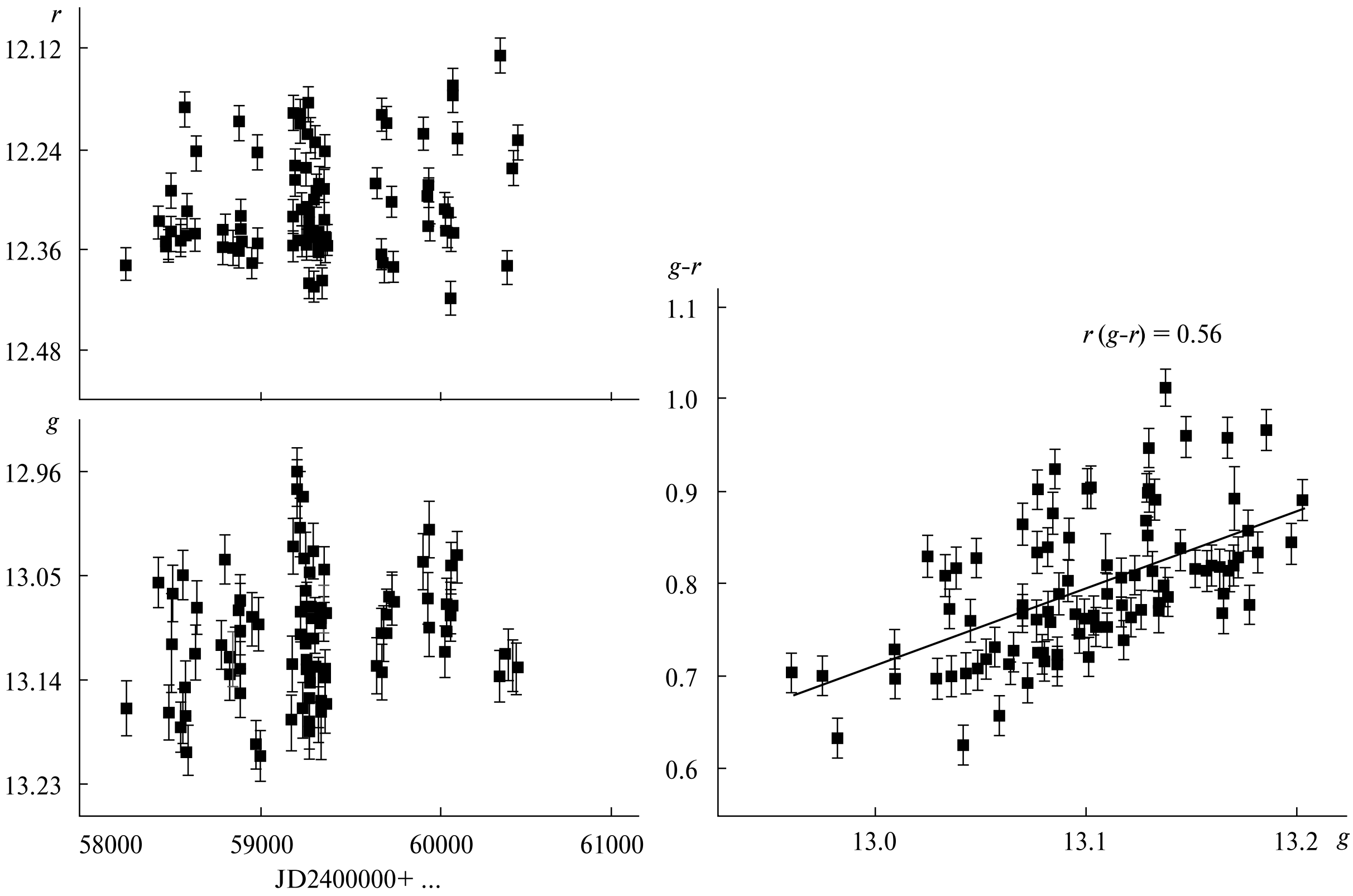

***Figure*** 9. NGC 3521. The light curves (left) and $g - r$ colour (right) indices according to the data of ZTF observations in *gr*-filters for 2018—2024

*Table 3.* **NGC 3521. The amplitude of variability in *BVR*-filters by the observational data in 2021—2022 with the Zeiss-600 telescope at the Terskol observatory**

| Object | $\Delta B$ | $Er\pm$ | $\Delta V$ | $Er\pm$ | $\Delta R$ | $Er\pm$ |
|---|---|---|---|---|---|---|
| NGC3521 | 0.116 | 0.003 | 0.079 | 0.088 | 0.052 | 0.001 |

*Table 4.* **NGC 3521. The amplitude of variability in *gr*-filters by the observational data with ZTF in 2018—2024**

| Object | $\Delta g$ | $Er\pm$ | $\Delta r$ | $Er\pm$ |
|---|---|---|---|---|
| NGC 3521 | 0.17 | 0.023 | 0.20 | 0.02 |

***Zwicky Transient Facility (ZTF) data***. We also found the available data from the Zwicky Transient Facility [38]. The light curves according to the data of ZTF observations in *gr*-filters for 2018—2024 are shown in Fig. 9 (left). NGC 3521 demonstrated variability with an amplitude of $0.17 \pm 0.023$ mag in the *g*-filter and $0.20 \pm 0.02$ mag in the *r*-filter (Table 4). The $g - r$ colour indices show a BWB trend (bluer-when-brighter) with a Pearson coefficient $r(g - r) = 0.56$ (Fig. 9, right), which is a medium correlation.

## 6. SUMMARY

We studied the multiwavelength properties of NGC 3521 from UV- to radio ranges exploiting the data from GALEX for UV-, SDSS for optical, 2MASS, WISE, MIPS (Spitster) and PACS, SPIRE (Herschel) for IR-, and NRAO VLA for radio ranges. To obtain the spectral energy distribution, we exploited the CIGALE software and constructed SEDs without (model A) and with (model B) AGN module. To confirm the type of nuclear activity of NGC 3521 as LINER, we used a BPT diagram.

We also present the results of the photometric data processing. Exploring the ZTF observations in 2018—2024, we found, for the first time, a weak photometric variability of the nuclear activity, where the correlation between $g - r$ colour indices and *g*-magnitude for long-term timescale shows a BWB trend (bluer-when-brighter) with a Pearson coefficient $r(g - r) = 0.56$, which is a medium correlation. We provided observations using a Zeiss-600 telescope with an aperture size of 8″ at the Terskol observatory im 2021—2022. To detect the variability of NGC 3521 during the day (IDV), we observed it in the *R*-filter with an exposure of 90 seconds for three hours on Feb 11, 2022. These data show a trend towards an increase in brightness with the amplitude of variability of $0.04 \pm 0.001$ mag.

According to the results of the simulations, the best fit to the observed SED is provided by model A, which considers the contribution to the radiation from all galaxy components, assuming that the galaxy nucleus (LINER type) is inactive. Within this model, we derived the stellar mass $M_{star} = 2.13 \times \times 10^{10}\ M_{Sun}$, the dust mass $M_{dust} = 8.45 \times 10^7\ M_{Sun}$, and the star formation rate $SFR = 1.76\ M_{Sun}$ /year, with the corresponding value of the statistic $\chi^2/d.o.f = 1.8$. Compared to the results obtained with model A, the model B with AGN module led to a slight increase in $\chi^2/d.o.f.$ The parameters obtained in both models were similar in value and order. These results are in the best agreement with the estimates obtained by other methods, for instance, by Leroy et al. [35], where the stellar mass $\lg(M_{star}/M_{Sun}) = 10.7$, the neutral hydrogen mass $\lg(M_{H\ I}/M_{Sun}) = 10.0$, and the star formation rate $SFR = 2.1\ M_{Sun}$ /year were determined.

Also, based on the radio data from the HIPASS catalogue by Allison et al. [1], we estimated the mass of neutral hydrogen to be $M_{H\ I} = 1.3 \times 10^{10}\ M_{Sun}$, which is of an order of magnitude larger than the mass of the stellar component. This result is comparable with the data by Elson [19], who used the THINGS and estimated $M_{H\ I} = 7.5 \times 10^9\ M_{Sun}$ and found a slow rotating halo H I component with $M_{H\ I} = 1.5\times 10^9\ M_{Sun}$.

Given that the UV, optical, and near-IR observation area covers only a part of the galaxy, this SED region may distort the overall picture. In this context, the results obtained by Warren et al. [64] for radio range do not contradict what we received. These authors did not find a correlation between the star formation efficiency and the gas surface density; however, they found that the star formation efficiency of the dense molecular gas is slightly declining as a function of molecular gas density. Zibetti et al. studied NGC 3521 among other nearby galaxies in the optical to the mid-IR range [66]. Constructing their SED on a pixel-by-pixel basis, they found a disconnect between the optical and IR when normalized to the near-IR (H band) data. However, due to the complex dust geometry and high position angle of NGC 3521, these authors could not find a correlation of variations between optical- and IR-dominated components discovered for other studied galaxies but confirmed the well-known optical-IR colour correlations: emission from stars contributes to the optical range and absorbed by dust, which emits in the IR.

The next stage of our work will be to obtain precise spectrophotometric data for apertures sufficient to cover the entire visible part of the galaxy from UV to radio, including decameter, ranges [33]. Preliminary search through available catalogues and sky survey [57] shows that there is a significant lack of data to construct SEDs of the Milky Way galaxies-analogies with a complete set of their multiwavelength properties. But our results of NGC 3521 altogether with the multiwavelength data analysis of other isolated galaxies with a weak nuclear activity can help in making decision to use SEDs as the MWA indicators.

***Acknowledgements***. O. Kompaniiets expresses her gratitude to Prof. Agnieszka Pollo and Prof. Katarzyna Małek for the opportunity to do the internship at the National Centre for Nuclear Research (Poland) for mastering basic techniques to work with the CIGALE software. Vavilova I. B. notes that her work is supported by the National Research Fund of Ukraine (Project No. 2023.03/0188). The research on the properties of isolated galaxies provided by Kompaniiets O. V., Izviekova I. O., and Vavilova I. B. is a part of the project, which received funding through the EURIZON project, which is funded by the European Union under grant agreement No.871072.

This research has made use of the NASA/IPAC Extragalactic Database (NED), which is funded by NASA and operated by the California Institute of Technology. This research has used the «Aladin sky atlas» developed at CDS, Strasbourg Observatory, France.

Part of this research is based on observations from the Zwicky Transient Facility project which are available in data release 19. Based on observations obtained with the 48-inch Samuel Oschin Telescope at the Palomar Observatory as part of the Zwicky Transient Facility project. Based on observations obtained with the Samuel Oschin Telescope 48-inch Palomar Observatory as part of the Zwicky Transient Facility project. ZTF is supported by the National Science Foundation under Grants No. AST-1440341 and AST-2034437 and a collaboration including current partners Caltech, IPAC, the Oskar Klein Center at Stockholm University, the University of Maryland, University of California, Berkeley , the University of Wisconsin at Milwaukee, University of Warwick, Ruhr University, Cornell University, Northwestern University and Drexel University. Operations are conducted by COO, IPAC, and UW.

Follow-up data were uploaded and calibrated using the Black Hole Target Observation Manager (BHTOM)3 tool for coordinated observations and processing of photometric time series. The BHTOM project has received funding from the European Union's Horizon 2020 research and innovation program under grant agreement No. 101004719 (OPTICON-RadioNet Pilot, ORP). BHTOM acknowledges the following people who helped with its development: Patrik Sivak, Kacper Raciborski, Piotr Trzcionkowski, and AKOND company.

REFERENCES

1. Allison, J. R., Sadler, E. M., Meekin, A. M. (2014). A search for H I absorption in nearby radio galaxies using HIPASS. *Mon. Notic. Roy. Astron. Soc.*, **440**, Is. 1, 696—718. DOI: 10.1093/mnras/stu289, arXiv: 1402.3530 [astro-ph.GA].
2. Baldwin, J. A., Phillips, M. M., Terlevich, R. (1981). Classification parameters for the emission-line spectra of extragalactic objects. *Publ. Astron. Soc. Pacif.*, **93**, 5—19. DOI: 10.1086/130766.
3. Bonnarel, F., Fernique, P., Bienaymé, O. et al. (2000). The ALADIN interactive sky atlas. A reference tool for identification of astronomical sources. *Astron. and Astrophys. Suppl. Ser.*, **143**, 33—40.
4. Boquien, M., Burgarella, D., Roehlly, Y., et al. (2019). CIGALE: a python Code Investigating GALaxy Emission. *Astron. and Astrophys.*, **622**, A103. DOI: 10.1051/0004-6361/201834156, arXiv: 1811.03094 [astro-ph.GA].
5. Brown, M. J. I., Armus, L., Calvin, D. E., et al. (2014). An Atlas of Galaxy Spectral Energy Distributions from the Ultraviolet to the Mid-infrared. *Astrophys. J. Suppl. Ser.*, **212**, Is. 2, 18. DOI: 10.1088/0067-0049/212/2/18, arXiv: 1312.3029 [astro-ph.CO].
6. Bruzual, G., Charlot, S. (2003). Stellar population synthesis at the resolution of 2003. *Mon. Notic. Roy. Astron. Soc.*, **344**, Is. 4, 1000—1028. DOI: 10.1046/j.1365-8711.2003.06897.x, arXiv: astro-ph/0309134 [astro-ph].
7. Burbidge, E. M., Burbidge, G. R., et al. (1964). The Rotation and Mass of NGC 3521. *Astrophys. J.*, **139**, 1058. DOI: 10.1086/147845.
8. Calzetti, D., Armus, L., Bohlin, R. C., et al. (2000). The Dust Content and Opacity of Actively Star-forming Galaxies. *Astrophys. J.*, **533**, Is. 2, 682—695. DOI: 10.1086/308692, arXiv: astro-ph/9911459 [astro-ph].
9. Casertano, S., van Gorkom, J. H. (1991). Declining rotation curves — The end of a conspiracy? *Astron. J.*, **101**, 1231—1241.
10. Chabrier, G. (2003). Galactic Stellar and Substellar Initial Mass Function. *Publ. Astron. Soc. Pacif.*, **115**, Is. 809, 763—795. DOI: 10.1086/376392, arXiv: astro-ph/0304382 [astro-ph].
11. Ciesla, L., Elbaz, D., Fensch, J. (2017). The SFR-M* main sequence archetypal star-formation history and analytical models. *Astron. and Astrophys.*, **608**, A41. DOI: 10.1051/0004-6361/201731036, arXiv: 1706.08531 [astro-ph.GA].
12. Coccato, L., Fabricius, M., Saglia, R. P., et al. (2018). Spectroscopic decomposition of NGC 3521: unveiling the properties of the bulge and disc. *Mon. Notic. Roy. Astron. Soc.*, **477**, Is. 2, 1958—1969. DOI: 10.1093/mnras/sty705.
13. Condon, J. J., Cotton, W. D., Broderick, J. J. (2002). Radio Sources and Star Formation in the Local Universe. *Astrophys. J.*, **124**, Is. 2, 675—689. DOI: 10.1086/341650.
14. Dale, D. A., Aniano, G., Engelbracht, C. W., et al. (2012). Herschel Far-infrared and Submillimeter Photometry for the KINGFISH Sample of nearby Galaxies. *Astrophys. J.*, **745**, Is. 1, 95. DOI: 10.1088/0004-637X/745/1/95, arXiv: 1112.1093 [astro-ph.CO].
15. Das, M., Teuben, P. J., Vogel, S. N., et al. (2003). Central Mass Concentration and Bar Dissolution in Nearby Spiral Galaxies. *Astrophys. J.*, **582**, Is. 1, 190—195. DOI: 10.1086/344480.
16. Dettmar, R.-J., Skiff, B. A. (1993). Declining rotation curves in interacting galaxies. NASA Ames Research Center, Evolution of Galaxies and their Environment, 251—252.
17. Dobrycheva, D. V., Vavilova, I. B., Melnyk, O. V., Elyiv, A. A. (2018). Morphological type and color indices of the SDSS DR9 galaxies at $0.02 < z < 0.06$. *Kinemat. Phys. Celest. Bodies*, **34**, Is. 6, 290. doi: 10.3103/S0884591318060028
18. Draine, B. T., Dale, D. A., Bendo, G., et al. (2014). Andromeda's Dust. *Astrophys. J.*, **780**, Is. 2, 172. DOI: 10.1088/0004-637X/780/2/172, arXiv: 1306.2304 [astro-ph.CO].
19. Elson, E. C. (2014). An H I study of NGC 3521 — a galaxy with a slow-rotating halo. *Mon. Notic. Roy. Astron. Soc.*, **437**, Is. 4, 3736–3749. DOI: 10.1093/mnras/stt2182.
20. Fabricius, M. H., Coccato, L., Bender, R., et al. (2015). Regrowth of stellar disks in mature galaxies: The two component nature of NGC 7217 revisited with VIRUS-W. *IAU Proc.*, **309**, 81—84. DOI: 10.1017/S1743921314009363.
21. Fritz, J., Franceschini, A., Hatziminaoglou, E. (2006). Revisiting the infrared spectra of active galactic nuclei with a new torus emission model. *Mon. Notic. Roy. Astron. Soc.*, **366**, Is. 3, 767—786. DOI: 10.1111/j.1365-2966.2006.09866.x.
22. Grasha, K., Chen, Q. H., Battisti, A. J., et al. (2022). Metallicity, Ionization Parameter, and Pressure Variations of H II Regions in the TYPHOON Spiral Galaxies: NGC 1566, NGC 2835, NGC 3521, NGC 5068, NGC 5236, and NGC 7793. *Astrophys. J.*, **929**, Is. 2, 118. DOI: 10.3847/1538-4357/ac5ab2, arXiv: 2203.02522 [astro-ph.GA].
23. Haslbauer, M., Banik, I., Kroupa, P., et al. (2024). The Magellanic Clouds are very rare in the IllustrisTNG simulations. *Universe*, **10**(10), 385. DOI: 10.3390/universe10100385.
24. Heida, M., Jonker, P. G., Torres, M. A. P., et al. (2014). Near-infrared counterparts of ultraluminous X-ray sources. *Mon. Notic. Roy. Astron. Soc.*, **442**, Is. 2, 1054—1067. DOI: 10.1093/mnras/stu928.
25. Helou, G., Soifer, B. T., Rowan-Robinson, M. (1985). Thermal infrared and nonthermal radio: remarkable correlation in disks of galaxies. *Astrophys. J.*, **298**, L7—L11. DOI: 10.1086/184556.

26. Inoue, A. K. (2011). Rest-frame ultraviolet-to-optical spectral characteristics of extremely metal-poor and metal-free galaxies. *Mon. Notic. Roy. Astron. Soc.*, **415**, Is. 3, 2920—2931. DOI: 10.1111/j.1365-2966.2011.18906.x, arXiv: 1102.5150 [astro-ph.CO].
27. Karachentsev, I. D., Makarova, L. N., Anand, G. S., et al. (2022). Around the Spindle Galaxy: The Dark Halo Mass of NGC 3115. *Astron. J.*, **163**, Is. 5, 234. DOI: 10.3847/1538-3881/ac5ab5.
28. Karachentseva, V. E. (1973). The Catalogue of Isolated Galaxies. *Astrof. Issledovanija Byu. Spec. Ast. Obser.*, **8**, 3—49.
29. Kauffmann, G., Heckman, T. M., Tremonti, C., et al. (2003). The host galaxies of active galactic nuclei. *Mon. Notic. Roy. Astron. Soc.*, **346**, Is. 4, 1055—1077. DOI: 10.1111/j.1365-2966.2003.07154.x, arXiv: astro-ph/0304239 [astro-ph].
30. Kewley, L. J., Dopita, M. A., Sutherland, R. S., et al. (2001). Theoretical Modeling of Starburst Galaxies. *Astrophys. J.*, **556**, Is. 1, 121—140. DOI: 10.1086/321545, arXiv: astro-ph/0106324 [astro-ph].
31. Knapik, J., Soida, M., Dettmar, R.-J., et al. (2000). Detection of spiral magnetic fields in two flocculent galaxies. *Astron. and Astrophys.*, **362**, 910—920. DOI: 10.48550/arXiv.astro-ph/0009438.
32. Kompaniiets, O. V. (2023). Multiwavelength properties of the low-redshift isolated galaxies with active nuclei modelled with CIGALE. *Space Sci. and Technol.*, **29**(5), 88—98. DOI: 10.15407/knit2023.05.088.
33. Konovalenko, A., Sodin, L., Zakharenko, V., et al. (2016). The modern radio astronomy network in Ukraine: UTR-2, URAN and GURT. *Exp. Astron.*, **42**, Is. 1, 11—48. doi: 10.1007/s10686-016-9498-x
34. Kourkchi, E., Tully, R. B. (2017). Galaxy Groups Within 3500 km/s. *Astrophys. J.*, **843**, Is. 1, 16. DOI: 10.3847/1538-4357/aa76db.
35. Leroy, A. K., Walter, F., Bigiel, F., et al. (2008). The Star Formation Efficiency in Nearby Galaxies: Measuring Where Gas Forms Stars Effectively. *Astrophys. J.*, **136**, Is. 6, 2782—2845. DOI: 10.1088/0004-6256/136/6/2782, arXiv: 0810.2556 [astro-ph].
36. Liu, G., Koda, J., Calzetti, D., et al. (2011). The Super-linear Slope of the Spatially Resolved Star Formation Law in NGC 3521 and NGC 5194 (M51a). *Astrophys. J.*, **735**, Is. 1, 63. DOI: 10.1088/0004-637X/735/1/63.
37. López, K. M., Jonker, P. G., Heida, M. (2015). Discovery and analysis of a ULX nebula in NGC 3521. *Mon. Notic. Roy. Astron. Soc.*, **489**, Is. 1, 1249—1264. DOI: 10.1093/mnras/stz2127.
38. Masci, F. J., Laher, R. R., Rusholme, B., et al. (2018). The Zwicky Transient Facility: Data Processing, Products, and Archive. *Publ. Astron. Soc. Pacif.*, **131**, 995.
39. McGaugh, S. S. (2016). The Surface Density Profile of the Galactic Disk from the Terminal Velocity Curve. *Astrophys. J.*, **816**, Is. 1, 42. DOI: 10.3847/0004-637X/816/1/42.
40. Melnyk, O., Karachentseva, V., Karachentsev, I. (2015). Star formation rates in isolated galaxies selected from the Two-Micron All-Sky Survey. *Mon. Notic. Roy. Astron. Soc.*, **451**, Is. 2, 14. DOI: 10.1093/mnras/stv950
41. Moustakas, J., Kennicutt, R. C. J. (2006). An Integrated Spectrophotometric Survey of Nearby Star-forming Galaxies. *Astrophys. J. Suppl. Ser.*, **164**, Is. 1, 81—98. DOI: 10.1086/500971, arXiv: astro-ph/0511729 [astro-ph].
42. Moustakas, J., Kennicutt, R. C. J. (2007). VizieR Online Data Catalog: Spectrophotometry of nearby galaxies (Moustakas+, 2006). *VizieR Online Data Catalog*, *J/ApJS/**164**/81.* DOI: 10.26093/cds/vizier.21640081.
43. Muñoz-Mateos, J. C., Gil de Paz, A., Boissier, S., et al. (2009). Radial Distribution of Stars, Gas, and Dust in SINGS Galaxies. I. Surface Photometry and Morphology. *Astrophys. J.*, **703**, Is. 2, 1569—1596. DOI: 10.1088/0004-637X/703/2/1569, arXiv: 0909.2648 [astro-ph.CO].
44. Netzer, H. (1990). AGN emission. *Active Galactic Nuclei*. Eds R. D. Blandford, H. Netzer, L. Woltjer.
45. Noll, S., Burgarella, D., Giovannoli, E., et al. (2009). Analysis of galaxy spectral energy distributions from far-UV to far-IR with CIGALE: studying a SINGS test sample. *Astron. and Astrophys.*, **507**, Is. 3, 1793—1813. DOI: 10.1051/0004-6361/200912497, arXiv: 0909.5439 [astro-ph.CO].
46. Pilyugin, L. S., Tautvaišienė, G., Lara-López, M. A. (2023). Searching for Milky Way twins: Radial abundance distribution as a strict criterion. *Astron. and Astrophys.*, **676**, A57. DOI: 10.1051/0004-6361/202346503, arXiv: 2306.09854 [astro-ph.GA].
47. Pulatova, N. G., Vavilova, I. B., Sawangwit, U., et al. (2015). The 2MIG isolated AGNs. I. General and multiwavelength properties of AGNs and host galaxies in the northern sky. *Mon. Notic. Roy. Astron. Soc.*, **447**, Is. 3, 2209—2223. DOI: 10.1093/mnras/stu2556
48. Pulatova, N. G., Vavilova, I. B., Vasylenko, A. A., et al. (2023). Radio properties of the low-redshift isolated galaxies with active nuclei. *Kinemat. Phys. Celest. Bodies*, **39**, Is. 2, 47—72. DOI: 10.15407/kfnt2023.02.047
49. Regan, M. W., Thornley, M. D., Vogel, S. N., et al. (2006). The Radial Distribution of the Interstellar Medium in Disk Galaxies: Evidence for Secular Evolution. *Astrophys. J.*, **652**, Is. 2, 1112—1121. DOI: 10.1086/505382.
50. Rosolowsky, E., Hughes, A., Leroy, A., et al. (2021). Giant molecular cloud catalogues for PHANGS-ALMA: methods and initial results. *Mon. Notic. Roy. Astron. Soc.*, **502**, Is. 1, 1218—1245. DOI: 10.1093/mnras/stab085.

51. Schawinski, K., Kaviraj, S., Khochfar, S., et al. (2007). Observational evidence for AGN feedback in early-type galaxies. *Mon. Notic. Roy. Astron. Soc.*, **382**, Is. 4, 1415—1431. DOI: 10.1111/j.1365-2966.2007.12487.x, arXiv: 0709.3015 [astro-ph].
52. Serote Roos, M., Boisson, C., Joly, M. (1998). Stellar populations in active galactic nuclei - I. The observations. *Mon. Notic. Roy. Astron. Soc.*, **301**, Is. 1, 1—14. DOI: 10.1046/j.1365-8711.1998.01462.x, arXiv: astro-ph/9804033 [astro-ph].
53. Spinoglio, L., Malkan, M. A. (1989). The 12 Micron Galaxy Sample. I. Luminosity Functions and a New Complete Active Galaxy Sample. *Astrophys. J.*, **342**, 83. DOI: 10.1086/167577.
54. Thornley, M. D. (1996). Uncovering Spiral Structure in Flocculent Galaxies. *Astrophys. J.*, **469**, 45. DOI: 10.1086/310250.
55. Van Gorkom, J. H., Knapp, G. R., Ekers, R. D., et al. (1986). The distribution and kinematics of H I in the active elliptical galaxy NGC 1052. *Astrophys. J.*, **91**, 791—807. DOI: 10.1086/114060.
56. Vasylenko, A. A., Vavilova, I. B., Pulatova, N. G. (2020). Isolated AGNs NGC 5347, ESO 438-009, MCG-02-04-090, and J11366-6002: Swift and NuSTAR joined view1. *Astron. Nachr.*, **341**, Is. 8, 801—811. DOI: 10.1002/asna.202013783
57. Vavilova, I., Dobrycheva, D., Vasylenko, M. et al. (2020). Multiwavelength extragalactic surveys: Examples of data mining. Knowledge discovery in big data from astronomy and Earth observation, 1st Edition. Ed. by P. Skoda and A. Fathalrahman. Elsevier, 307-323. doi: 10.1016/B978-0-12-819154-5.00028-X
58. Vavilova, I. B., Dobrycheva, D. V., Vasylenko, M. Y., et al. (2021). Machine learning technique for morphological classification of galaxies from the SDSS. I. Photometry-based approach. *Astron. and Astrophys.*, **648**, id. A122, 14 p. DOI: 10.1051/0004-6361/202038981
59. Vavilova, I. B., Fedorov, P. N., Dobrycheva, D. V. et al. (2024). An advanced approach for definition of the "Milky Way galaxies-analogues". *Space Sci. and Technol.*, **30**(4), 81—90. DOI: 10.15407/knit2024.04.081.
60. Vavilova, I. B., Khramtsov, V., Dobrycheva, D. V., et al. (2022), Machine learning technique for morphological classification of galaxies from SDSS. II. The image-based morphological catalogs of galaxies at $0.02 < z < 0.1$. *Space Sci. and Technol.*, **28**, Is. 1, 3—22. Doi: 10.15407/knit2022.01.003
61. Vila-Vilaro, B., Cepa, J., Zabludoff, A. (2015). The Arizona Radio Observatory Survey of Molecular Gas in Nearby Normal Spiral Galaxies I: The Data. *Astrophys. J. Suppl. Ser.*, **218**, Is. 2, 28. DOI: 10.1088/0067-0049/218/2/28.
62. Vol'vach, A. E., Vol'vach, L. N., Kut'kin, A. M., et al. (2011). Multi-frequency studies of the non-stationary radiation of the blazar 3C 454.3. *Astron. Rep.*, **55**, Is. 7, 608—615. doi: 10.1134/S1063772911070092
63. Walter, F., Brinks, E., de Blok, W. J. G., et al. (2008). THINGS: The H I Nearby Galaxy Survey. *Astrophys. J.*, **136**, Is. 6, 2563—2647. DOI: 10.1088/0004-6256/136/6/2563, arXiv: 0810.2125 [astro-ph].
64. Warren, B. E., Wilson, C. D., Israel, F. P., et al. (2010). The James Clerk Maxwell Telescope Nearby Galaxies Legacy Survey. II. Warm Molecular Gas and Star Formation in Three Field Spiral Galaxies. *Astrophys. J.*, **714**, Is. 1, 571—588. DOI: 10.1088/0004-637X/714/1/571.
65. Zeilinger, W. W., Vega Beltrán, J. C., Rozas, M., et al. (2001). NGC 3521: Stellar Counter-Rotation Induced by a Bar Component. *Astrophys. and Space Sci.*, **276**, Is. 2/4, 643—650. DOI: 10.1023/A:1017548101623.
66. Zibetti, S., Groves, B. (2011). Resolved optical-infrared spectral energy distributions of galaxies: universal relations and their breakdown on local scales. *Mon. Notic. Roy. Astron. Soc.*, **417**, Is. 2, 812—834. DOI: 10.1111/j.1365-2966.2011.19286.x

*О. С. Пастовень*[1,2], пров. інж.
https://orcid.org/0009-0000-4309-034X
*О. В. Компанієць*[1], мол. наук. співроб.
https://orcid.org/0000-0002-8184-6520
E-mail: kompaniets@mao.kiev.ua
*І. В. Вавилова*[1], д-р фіз.-мат. наук, проф., зав. відділу
https://orcid.org/0000-0002-5343-1408
*І. О. Ізвєкова*[1,3], мол. наук. співроб.
https://orcid.org/0009-0009-4307-0627

[1]Головна астрономічна обсерваторія Національної академії наук України
вул. Академіка Заболотного 27, Київ, Україна, 03143
[2] Фізичний факультет Київського національного університету імені Тараса Шевченка
проспект Академіка Глушкова 4, Київ, Україна, 03680
[3] Міжнародний центр астрономічних та медико-екологічних досліджень
вул. Академіка Заболотного 27, Київ, Україна, 03143

## NGC 3521 ЯК АНАЛОГ МОЛОЧНОГО ШЛЯХУ: СПЕКТРАЛЬНИЙ РОЗПОДІЛ ЕНЕРГІЇ ВІД УФ- ДО РАДІОДІАПАЗОНУ І ФОТОМЕТРИЧНА ЗМІННІСТЬ

Досліджено мультихвильові властивості NGC 3521, галактики-близнюка Молочного Шляху, від ультрафіолетового до радіодіапазону, з використанням даних GALEX для ультрафіолетового, SDSS для оптичного, 2MASS, WISE, MIPS (Spitzer) і PACS, SPIRE (Herschel) для інфрачервоного та NRAO VLA для радіодіапазону. Для побудови спостережного спектрального розподілу енергії (SED) було використано програмне забезпечення CIGALE, у якому величину SED було змодельовано із використанням моделі A без урахування активного ядра і моделі B з модулем активного ядра галактики. Для підтвердження типу активності ядра NGC 3521 як LINER побудовано BPT-діаграму.

Досліджуючи дані спостережень ZTF для NGC 3521 у 2018—2024 роках, ми вперше виявили, що показники кольору $g - r$ мають тренд BWB (що яскравіший, то синіший) із коефіцієнтом Пірсона $r(g - r) = 0.56$, що є середнім ступенем кореляції. Додатково у роботі наведено результати обробки фотометричних даних, отриманих нами на телескопі Zeiss-600 з апертурою 8″ в обсерваторії Терскол у 2021—2022 роках. Для виявлення змінності NGC 3521 протягом дня (IDV) було проведено спостереження у R-смузі з експозицією 90 с протягом трьох годин 11 лютого 2022 року. Дані демонструють тенденцію до збільшення блиску з амплітудою змінності $0.04 \pm 0.001^m$.

За результатами моделювання SED спостережуваним даним найкраще відповідає модель A, яка враховує внесок у випромінювання всіх компонентів галактики, припускаючи, що ядро галактики неактивне. У рамках моделі A отримано масу зоряного компонента $M_{star} = 2.13\cdot10^{10}\ M_{Sun}$, масу пилу $M_{dust} = 8.45\cdot10^{7}\ M_{Sun}$ та темп зореутворення $SFR = 1.76\ M_{Sun}$/рік з відповідним значенням $\chi^2$/d.o.f = 1.8. На основі радіоданих з каталогу HIPASS було оцінено також масу нейтрального водню $M_{\mathrm{H\ I}} = 1.3\cdot10^{10}\ M_{Sun}$, що за порядком збігається з масою зоряного компонента.